# GALACTIC COSMIC RAYS – CLOUDS EFFECT AND BIFURCATION MODEL OF THE EARTH GLOBAL CLIMATE. PART 1. THEORY

[1]*Vitaliy D. Rusov* [a,bc], *Alexandr V. Glushkov* [d], *Vladimir N. Vaschenko* [c],

*Oksana T. Mykhalus* [a], *Yuriy A. Bondartchuk* [a], *Vladimir P. Smolyar* [a],

*Elena P. Linnik* [a], *Strachimir Cht. Mavrodiev* [e], *Boyko I. Vachev* [e]

[a] Odessa National Polytechnic University, Shevchenko av. 1, Odessa, 65044 Ukraine

[b] Bielefeld University, 25, University Str., Bielefeld, 33615 Germany

[c] Ukrainian National Antarctic Center, 16, Tarasa Schevchenko Blvd., Kiev, 01601 Ukraine

[d] Odessa State Environmental University, Odessa, Ukraine, 15, Lvivska str.,Odessa, 65105 Ukraine

[e] Institute for Nuclear Research and Nuclear Energy, Bulgarian Academy of Sciences, 72, Tsarigradsko shaussee blvd. 1784, Sofia, Bulgaria

## Abstract

Based on theoretical and experimental consideration of the first (the Twomey effect) and second indirect aerosol effects the quasianalytic description of physical connection between the galactic cosmic rays intensity and the Earth's cloud cover is obtained.

It is shown that the basic equation of the Earth's climate energy-balance model is described by the bifurcation equation (with respect to the temperature of the Earth's surface) in the form of assembly-type catastrophe with the two governing parameters defining the variations of insolation and Earth's magnetic field (or the galactic cosmic rays intensity in the atmosphere), respectively.

The principle of hierarchical climatic models construction, which consists in the structural invariance of balance equations of these models evolving on different time scales, is described.

---

[1] Corresponding author: Prof. Rusov V.D., Head of Department of Theoretical and Experimental Nuclear Physics, Odessa National Polytechnic University, 1, Shevchenko ave., Odessa, 65044, Ukraine,
Tel.: +380 487 348 556 , Fax:+380 482 641 672, E-mail: siiis@te.net.ua

# 1. Introduction

In aerosol physics of cloud formation  there are now profound and insuperable, at the first sight (Carslaw, 2009) contradictions caused by the anomalous low efficiency of "ion-aerosol clear-air mechanism" for nucleation of the sufficient number of condensation nuclei in an atmosphere (Carslaw, 2009; Pierce and Adams, 2009; Sloan and Wolfendale, 2008; Erlykin et al, 2008). In spite of this, many researchers have serious reasons (Carslaw, 2009) to consider that galactic cosmic rays (GCR) play one of key roles in the mechanisms of the weather and climate formation at our planet (Carslaw, 2009; Marsh and Svenmark, 2000; Svensmark and Calder, 2007; Stozhkov, 2003; Usoskin et al, 2004; Shaviv, 2002, 2003, 2003a).

One of primary purposes of this paper is development of such a model of the Earth's global climate, within the framework of which above mentioned contradictions are overcame by the simple GCR-clouds effect submodel, which takes into account interdependent micro- and macrophysics of so-called indirect aerosol effects (Twomey, 1977, 1991; Kaufman and Fraser, 1997; Breon, 2002; Kaufman, 2002; Lyon, 2000). In this connection, we will do the following digression.

Summarizing the outcomes of numerous studies concerned with the influence of GCR on atmospheric processes, particularly, on charged aerosol formation (the condensation centres of main greenhouse gas, i.e., water vapour) the following causal sequence of events can be appointed (Marsh and Svenmark, 2000; Svensmark and Calder, 2007; Stozhkov, 2003; Usoskin et al, 2004; Shaviv, 2002, 2003, 2003a, 2006; Lyon, 2000): the radiating sun $\rightarrow$ the variations of solar activity and insolation $\rightarrow$ modulation of GCR flux $\rightarrow$ cloud cover and thunderstorm activity variations $\rightarrow$ albedo variations $\rightarrow$ weather and climate changes.

Macroscopic physics of GCR flux modulation due to the solar wind is sufficiently evident. The main reason lies in the fact that the connection between the solar wind and the Earth's magnetosphere is indirect and controlled by the magnetic field of the solar wind through  magnetic reconnection (Priest and Forbes, 2000; Rind, 2002). At the same time, the GCR flux intensity in the lower atmosphere is modulated by long-term variations of the Earth's magnetic field caused by interaction of interplanetary magnetic field and magnetic perturbations due to solar activity. In practice long-term variations of the Earth's magnetic field are approximated by the vertical cutoff rigidity at different positions on the Earth, which are calculated in the framework of living models of geomagnetic field (Kudela and Bobik, 2004; Shea and Smart, 2003, 2004).

As far back as at the 1930-ies Forbush (1937) obtained for the first time the experimental data of strong inverse correlation between the cosmic ray intensity and solar activity (Ney, 1959). Now the great number of works is devoted to analogous data on GCR intensity. In particular, Bazilevskaya (2005) illustrated the explicit and exact example of inverse correlation between the GCR intensity with energy above 1.5 GeV and the solar activity (protons with energy above 1.0 GeV) on basis of experimental data during the 1958-2002 years. It should be added that the process of the GCR flux modulation at the time scales from some days (Forbush phenomenon) to the two tens of years (11-year solar cycle) is experimentally validated with a sufficient clarity, whereas the determination of GCR intensity variations, for example in the past on centennial and millennial timescales, is more intricate problem. This is caused by the fact that construction of complete theory of magnetic reconnection, which is a central problem in magnetospheric physics, is uncompleted until now in spite of some progress made in the last years (Priest and Forbes, 2000; Coley, 1985; Rind, 2002).

The close connection between the GCR intensity and global cloud cover was found by Svensmark and Friis-Christensen (1997). Their results demonstrate the strong positive correlation of GCR flux and cloud cover during a long-term cosmic rays modulation in the 11-year solar cycle. Furthermore, it was found, that the Earth's temperature depends more strongly on almost ten-year variations of GCR flux and solar cycle length, than on other solar activity parameters (Svensmark, 1998). Basing on these discoveries, Svensmark et al. (Marsh and Svensmark, 2000; Svensmark and Calder, 2007) made a strong conclusion that "…solar activity variations may be linked to climate variability by a chain "the solar wind - GCR - climate". Moreover, the fact that the influence of solar activity variation is strongest just at low cloudiness ($\leq 3$ km) was experimentally found. This in its turn allowed directly to point out the microphysical mechanism of such an influence including aerosol formation, which is enhanced due to ionization under cosmic rays action (Marsh and Svensmark, 2000; Svensmark and Calder, 2007).

It is known that aerosols play key part in the cloud formation and directly influence on the Earth radiation balance through the increase of albedo of the system "earth-atmosphere". In spite of the fact that their exact role is currently uncertain, in the first place such an effect is caused by a few "laws", which were revealed within the confines of atmospheric physics (the so-called aerosol indirect effect on clouds). For example, aerosols can also act as cloud condensation nuclei (CCN), increasing the number of droplets in clouds, which tend to decrease the average droplet size and may increase the cloud albedo (Twomey, 1977; Twomey, 1991), depending on the aerosol absorption and cloud optical thickness (Kaufman and Fraser, 1997). This process, called as the "Twomey effect" or the "first indirect" aerosol radiation forcing, renders cooling effect on climate (Breon et al, 2002). On the other hand, the high concentration of aerosols favours the new CCN formation causing the condensation of additional water vapour due to cloud cooling. At the same time, the small droplets have low probability to collide with each other and form precipitation. This change of "precipitation efficiency" predetermined by the increase of liquid water content in the cloud, lifetime of cloud and area of cover, is named the second indirect aerosol effect. The global importance of this effect is still not clear (Kaufman et al, 2002).

It is obvious that cosmic ray effect and the indirect aerosol effect are similar in that both drive the change of aerosol number (Carslaw et al, 2002). Therefore, in spite of a number of important distinctions denoted in Ref. (Carslaw et al, 2002), one can suppose with certainty that these effects are related to each other by deep and fine microphysics at the level of various, but possibly competing mechanisms of formation and growth of atmospheric aerosols (Fig.1 (Kulmala, 2003)). In this sense, until the theory of atmospheric aerosol formation can be fully understood, we can only hope that an empirical relation between the variations of cosmic ray intensity and the well-known experimental value of so-called "aerosol index" (AI) exists. Note, that the aerosol index characterizes the aerosol number in the atmospheric air column of unit cross-section, which is usually measured as the function of the average effective radius of cloud droplet (CDR) (Fig. 2), i.e., in the form CDR=$f$(AI) (Breon et al, 2002; Kaufman et al, 2002).

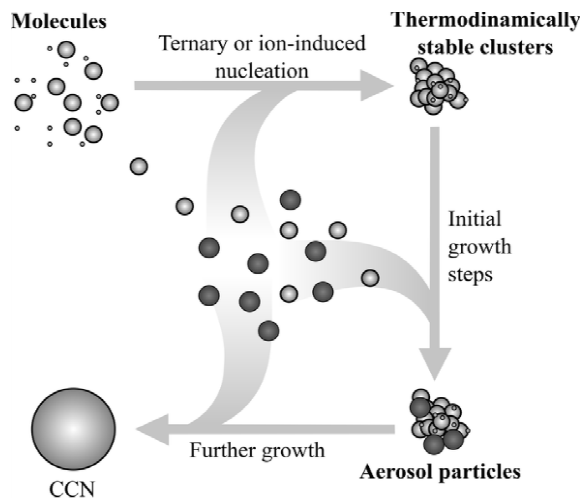

Fig. 1. Schematic sketch of the mechanism of neutral and ion-induced aerosol particle formation and growth (Kulmala, 2003).

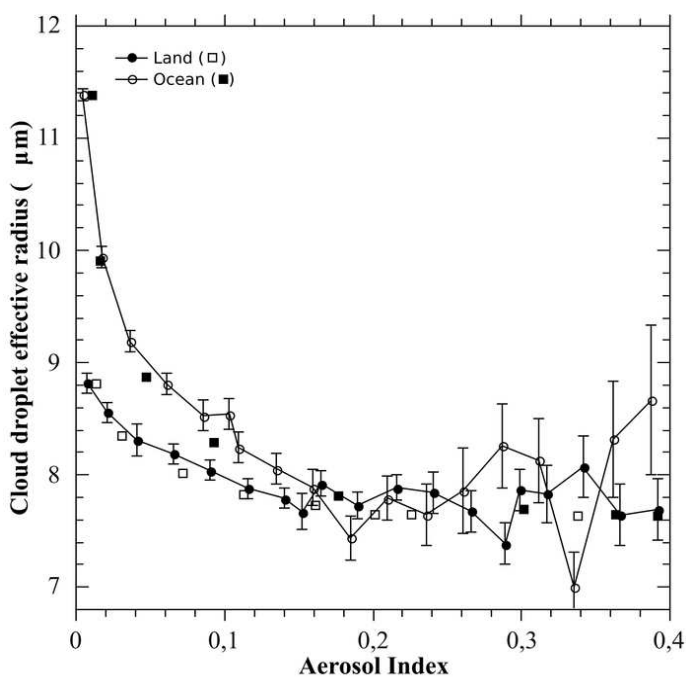

Fig. 2. The experimental dependence of the average cloud droplet effective radius (CDR) on aerosol index (AI) for a land (black circles, lower curve) and an ocean (white circles, upper curve) (Breon, 2002). Empiric values of CDR =$f$ (AI) ratio obtained by Eq.(1) are marked on the experimental curves for a land (■) and an ocean (□).

At the same time, it has to be noted, that researchers do not have clear understanding when the GCR modulation in above mentioned causal connection corresponds to the description of weather (stochastic by its nature) only, and when it corresponds to the stochastic or quasi-deterministic global climate. In our opinion, an answer to this question lies in the known spectrum of air temperature variations in the North Atlantic given by Kutzbach and Bryson (1974). The energy spectrum of variation periods shows (Fig. 3) that to the right of "window", i.e. deep and wide minimum, there is a spectrum of weather changes characterized by the white noise, whereas a spectrum of long-period climate fluctuations characterized by so called "red" noise, which differs from the white noise by a certain predictability (i.e. quasi-determinacy), lies to the left of this minimum. An intermediate spectrum of stochastic climate variations characterizing by the $1/f$ noise corresponds to the deep and wide minimum.

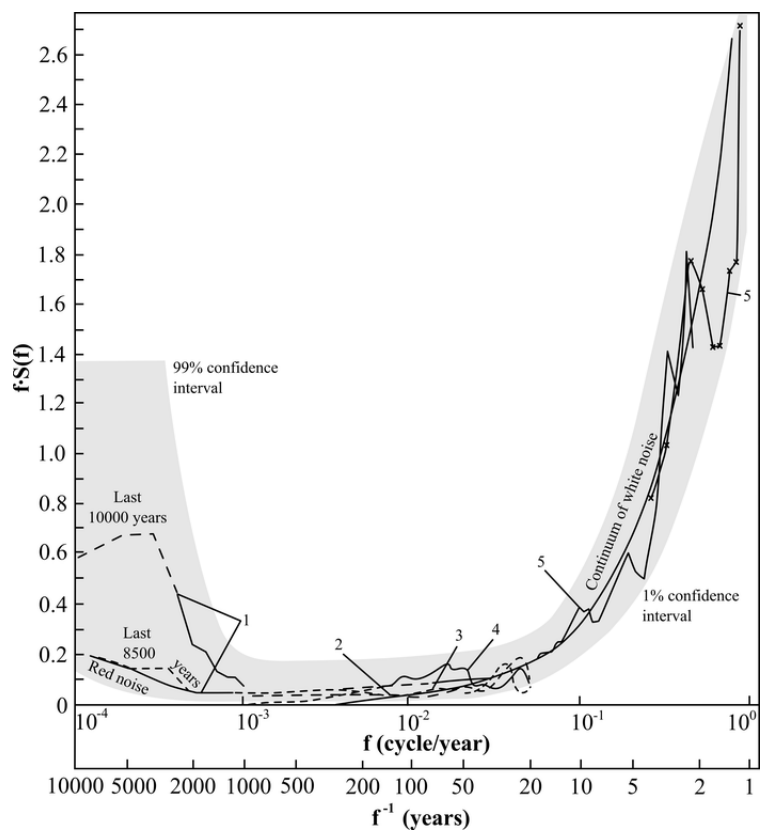

Fig. 3. Combined spectrum of the air temperature variations in the North-Atlantic sector of the terrestrial globe (Kutzbach and Bryson, 1974): $f$ are frequencies, cycle/year; $S(f)$ is spectral density; 1 – the Central England, paleobotany; 2 – the same, chronicles; 3 – the Iceland, chronicles; 4 – the Greenland, obtained by $\delta^{18}O$; 5 – the Central England, by Manley series (Kutzbach and Bryson, 1974).

Thus, the spectrum of temperature variations within the Earth climate system (ECS) shows that to describe the weather and climate[2] adequately the averaging of ECS parameters at appropriate time intervals is necessary. It is also obvious that the periods of ECS parameter averaging for stochastic description of weather and "intermediate" climate must be less than 30 years and 30-1000 years, respectively, whereas the determinate description of global climate evolution necessitates averaging periods ≥1000 years.

---

[2] Climate is commonly defined as the statistical ensemble of ECS states which is described by corresponding the set of parameters (temperature, pressure etc.) averaged over a long period of time. The classical period is 30 years (see fig.3), as defined by the World Meteorological Organization. From this it follows that the departure of weather from the climatic norm can not be examined as a change of climate.

It is surely ascertained (Yamasaki and Oda, 2002) that the spectral density of virtual axial dipole moment (VADM) fluctuations of the magnetic field of the Earth displays the variations with a period ~100 kyr, which correspond to the variations in the orbital eccentricity of the Earth, in the spectrum of paleointensity over the past 2.25 million years. Moreover, Yamasaki and Oda (2002) suggested that the geomagnetic field is, in some way, modulated by orbital eccentricity variations. In our opinion, this effect can be explained on phenomenological level, if to suppose that the gravitational field of the Sun (by way of indirect influence on the angular velocity of the Earth) affects convection processes (hydromagnetic dynamo) in the Earth liquid core and, thereby, "delegating" the variations to the Earth magnetic field with periods corresponding to periodicity of the Earth orbital eccentricity. Such a mechanism of the Earth precession as the cause of geomagnetism was for the first time proposed by Malkus (1968). Note that to explain a possible link between GCR and periodic changes of the Earth's orbital parameters Consolini and De Michelis (2003) have used the mechanism of such a kind in their work devoted to the role of stochastic resonance in the mechanism of geomagnetic pole reversals.

Thus, in spite of limited understanding the reason of appearance of periods typical for an excentricity in the magnetic field paleointensity spectrum , it is possible to conclude that just the millennial fluctuations of terrestrial magnetism intensity are peculiar modulator of GCR at the time scales >1000 years. It is also obvious that such a type of the GCR modulation directly relates to "deterministic" climate at the millennial time scale, and this fact can be used as a governing parameter in the models of global climate of the Earth.

This paper has for an object the development of an energy-balance model of climatic response to orbital variations, which takes into account an influence of galactic cosmic rays on global climate formation.

## 2. The Twomey aerosol effect and the temperature dependence of water volume (liquid and vapour) in the atmosphere

It is known, that the indirect observation of the Twomey effect (the first indirect aerosol effect) can be made by comparing the cloud droplet size and aerosol concentration. The dependence of CDR on AI (Fig. 2) was measured by special radiometers in real satellite observations (Breon et al, 2002; Kaufman et al, 2002) because CDR is more sensitive to the aerosol index than to the optical thickness. That is logical, so long as the aerosol index is a function of condensation nuclei concentration in the cloud (Breon et al, 2002).

It is easy to show that the experimental dependence of CDR on AI (Fig. 2) above an ocean and land can be represented (with approximation satisfactory for our purpose) as the following empirical dependence

$$AI = \left[ \frac{1}{(0.6r_{eff} - 4.385)r_{eff}} - \frac{\eta}{r_{eff}} \right]^{1.429}, \quad \eta = \begin{cases} 0, & ocean \\ 0.63, & land \end{cases} \tag{1}$$

where $r_{eff} = \langle r^3 \rangle / \langle r^2 \rangle$ is the average effective radius of the cloud droplet and $r$ is the radius of the cloud droplet.

We use below the fact that the number of aerosol particles $N_{CCN}$, which are CCN, and the number of cloud droplets $N_d$ are approximately connected by following expression (Twomey, 1977, 1977a, 1991):

$$N_d \cong (N_{CCN})^\alpha. \tag{2}$$

Taking into account that, on the one hand, cloud formation models and measurements indicate that $\alpha$ is of the order of 0.7 (Twomey, 1977, 1991; Rosenfeld et al, 2000), and, on the other hand, $AI \sim N_{CCN}$ (Twomey, 1977, 1991), using Eqs. (1)-(2) we can obtain the following expression for the concentration of cloud droplets:

$$N_d \approx \frac{1}{(0.6r_{eff} - 4.385)r_{eff}} - \frac{\eta}{r_{eff}}. \tag{3}$$

Then the volume $V_W$ of liquid water in the atmosphere "above an ocean" or "above a land" is equal to

$$V_w = p_\eta V_{atm} \cdot \frac{4}{3}\pi \langle r \rangle^3 N_d = p_\eta V_{atm} \frac{4\pi}{3} \frac{r_{eff}^2}{k_r^3} \left( \frac{1}{0.6r_{eff} - 4.385} - \eta \right), \tag{4}$$

where $V_{atm} \cong const$ is total volume of the atmosphere, $p_\eta$ is the part of the atmosphere volume "above ocean" or "above a land", $\langle r \rangle \approx k_r \cdot r_{eff}$ is the average radius of cloud droplet (CDR).

Therefore, the averaged total volume of liquid water in the atmosphere looks like

$$\langle V_w \rangle = V_{atm} \frac{4\pi}{3} \frac{r_{eff}^2}{k_r^3} \left[ \frac{1}{0.6r_{eff} - 4.385} - \frac{S_{land}\Lambda}{S_{ocean} + S_{land}} 0.63 \right], \tag{5}$$

$$\Lambda = \Lambda(r_{eff} - 7.7) - \Lambda(r_{eff} - 9.6)$$

where $\Lambda(x)$ is an integral of $\delta$-function:

$$\Lambda(x) = \int_{-\infty}^{x} \delta(z)dz = \begin{cases} 1, & x \geq 0, \\ 0, & x < 0, \end{cases}$$

$S_{land}$ and $S_{ocean}$ are the areas of land and oceans, respectively; $S_{land} / (S_{land} + S_{ocean}) \cong 0,29$.

It is noteworthy that the dependence (5) presented in Fig. 4 has clearly defined minimum at $r_{eff} \approx 14$ μm, which has, apparently, the physical sense of so-called precipitation threshold (Rosenfeld, 2000). Note that we will consider further the linear section of dependence (5) only, which is the left of minimum (Fig.4).

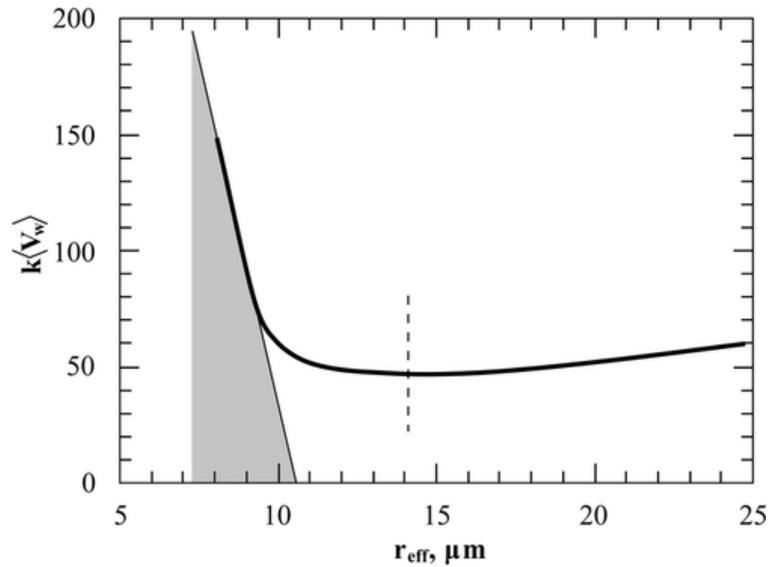

Fig. 4. The average total volume of liquid water $\langle V_w \rangle$ in the atmosphere as a function of cloud droplet effective radius (CDR). Minimum at $r_{eff} \approx 14$ μm (the vertical dotted line) corresponds to the so-called precipitation threshold.

Eq. (5) can be simplified since the real increments of temperature $\Delta T$ corresponding to the "warm" and "ice" ages of the Earth climate usually lie in a relatively narrow temperature range. Note that the range of increment of air average temperature on the millennial time scale experimentally obtained from EPICA Dome C (EPICA community members, 2004) and Vostok ice cores (Petit et al, 1999) are about $\Delta T = \pm 4 \div 6$ K. Taking under consideration this fact and using experimental data (Rosenfeld, 2000; Rosenfeld et al, 2002), it is easy to show that due to the inverse linear dependence the small increments of the average CDR, i.e. $\Delta r_{eff} \approx 2 \div 3$ μm, correspond to such increments $\Delta T$. Therefore, it can be supposed that by virtue of Eq. (3) the real scenarios of global climate "know" and "feel" only the relatively narrow range of the CDR values from a permitted values ($r_{eff} \approx 8 \div 14$ μm) located the left of the precipitation threshold line (Fig. 4). This property of climatic global scenarios makes it possible, in its turn, to simplify Eq. (5) and to write down it in the first approximation as the inverse linear dependence on CDR (see Fig. 4) or by virtue of experimental data (Rosenfeld, 2000; Rosenfeld et al, 2002) in the form of direct linear dependence on temperature:

$$\langle V_w \rangle \approx a - b r_{eff} = a_w + b_w T \ . \qquad (6)$$

Now, the following procedure to calculate the total water (vapour and liquid) in the atmosphere $\langle V_{w+v} \rangle$ can be offered. We have every reason to consider that the theoretical dependence of water vapour volume $\langle V_v \rangle$ on temperature is the same as Eq. (6) for water and differs from it in quantitative characteristics only. As regards this a whole number of serious hints is in scientific literature. For example, investigating the bistabillity of CCN concentrations and thermodynamics in the cloud-topped boundary layer, Baker and Charlson (1990) showed that the total water (vapour and liquid) and the temperature of CTBL are in the first approximation directly proportional to temperature of sea surface, above which this layer is. To a certain extent, it concerns the work of Albrecht (1989), where the model linear dependence of cloud volume on the sea surface temperature was obtained. This makes it possible to suppose that in the general case the total volume of water (vapour and liquid) in the atmosphere $\langle V_{w+v} \rangle$ is directly proportional to surface temperature, i.e.,

$$\langle V_{w+v} \rangle = \langle V_w \rangle + \langle V_v \rangle \sim T \,. \tag{7}$$

Therefore, if Eq. (6) is linear with respect to the temperature, in view of Eq. (7) the dependence of the total volume of water vapour $\langle V_v \rangle$ must also be linear with respect to the temperature. Then, taking into account stated above, the expression for the volume of water vapour $\langle V_v \rangle$ can be written with an allowance for (6) and (7) as follows

$$\langle V_v \rangle \approx a_v + b_v T \,. \tag{8}$$

As it mentioned above, likeness of the cosmic ray effect on clouds and the indirect aerosols effect manifests itself in the change of the number of aerosols (Carslaw et al, 2002). Although these effects, in our opinion, have common "microphysics", Eq. (1) does not contain a term responsible for the cosmic ray effect. In the next section, we try to examine the reasons for the absence of such a term and to "rehabilitate" it in the general case for various averaging intervals of ECS parameters.

The total volume of liquid water $V_w$, and water vapour $V_v$, in the atmosphere are calculated by Eq. (5). These calculations allow estimate the masses of $V_w$ and $V_v$ and, consequently, the re-emission energy characterizing both the total radiation energy of water $\Delta E_w$ and water vapour $\Delta E_v$ depending on the ECS temperature. Then, if an application of these terms in the energy-balance model of the Earth's climate will lead to the solution of energy-balance equation with respect to ECS temperature, and this solution will be in good agreement with experimental data of paleotemperature (e.g., the Vostok ice core data (Petit et al, 1999) and EPICA ice core data (EPICA community members, 2004)), our assumptions made to obtain Eq. (5) can be considered as acceptable (at least, until the direct experiments on research of the possible synergetic (with respect to insolation forcing) influence of GCR intensity on the aerosol index value in the form of dependence AI=$f$ (CDR) will be carried out).

Below we consider the energy-balance model of the Earth's global climate with two governing parameters.

## 3. Cosmic rays and energy-balance model of global climate

By virtue of energy conservation law, the real heat power of the Earth's radiation is approximately equal to the difference between the long-wave radiation power of Earth warmed-up surface $I(T,t)$ and heat energy power re-emitted by the liquid water $G_w(T,t)$, water vapour $G_v(T,t)$ and carbon dioxide $G_{CO_2}(T,t)$. For simplicity, we do not consider other greenhouse gases. Since the radiant equilibrium is reached during some tens of millenniums, an allowance for greenhouse effect results in the following energy-balance equations for the ECS (Fig. 5):

$$U(T,t) = P_{Sun}(t) \cdot [1 - \alpha(T)] - I_{Earth}(T) + \frac{1}{2} G_w(T,t) + \frac{1}{2} G_v(T,t) + \frac{1}{2} G_{CO_2}(T,t) \,, \tag{9}$$

where the left side of an equation $U(T, t)$, if it is nonzero, describes the so-called "inertial" power of heat variations in the ECS; $P_{Sun}(t)=(1/4(1-e^2))S_0 \cdot \gamma \approx (1/4)S_0 \cdot \gamma$ is the heat flow of solar radiation on the upper boundary of the atmosphere, W; $S_0=1366.2$ Wm$^{-2}$ is "solar constant" (Frohlich and Lean 1998); $e$ is the eccentricity of Earth elliptic orbit; $\gamma$ is the area of outer boundary of upper atmosphere, m$^2$; $\alpha$ is the ECS albedo; $I_{Earth}=\gamma\delta(\sigma T^4)$ is the gray radiation power of Earth surface, W; $\delta$=0,95 is coefficient of radiation gray chromaticity of Earth surface; $\sigma$=5,67·10$^{-8}$ is the Stephen-Boltzmann constant, Wm$^{-2}$K$^{-4}$; $T$ is the Earth surface temperature, K; $\gamma$ is the area of upper atmosphere outer boundary, $m^2$; $t$ is the time, for which the energy balance is considered.

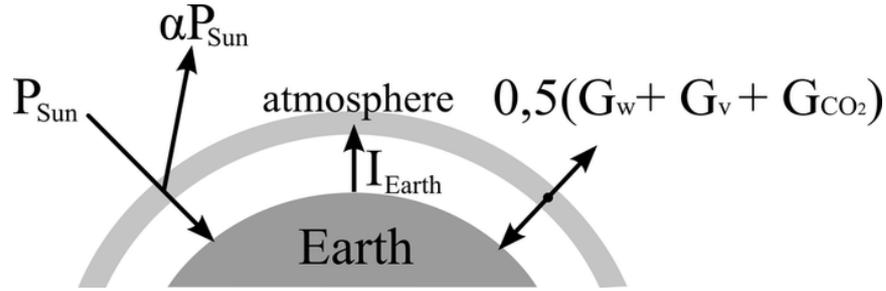

Fig. 5. The balance of energy flows on the Earth surface.

At first, we consider the functional dependence of heat energy power $G_w(T, t)$ re-emitted by liquid water on temperature. It is obvious that Eq. (6) for averaged total volume of liquid water in the atmosphere makes it possible to writing down the following expression for the re-emission power $G_w(T, t)$:

$$G_w(T,t) = \varepsilon_w \rho_w \theta \langle V_w \rangle, \qquad (10)$$

where $\varepsilon_w$ is the average radiation power per unit mass of liquid water, $\rho_w$ is the liquid water density; $\theta = \langle \Delta V_w \rangle / \langle V_w \rangle$ is the part of the near-surface liquid water volume in the clouds, which effectively reradiates the earlier absorbed (in the long-wave range) solar energy to the atmosphere. It is obvious, that due to the finite but small length of self-absorption of the reradiated energy in the clouds, the part of the effectively reradiated volume of liquid is inversely proportional to the total liquid water volume $<V_w>$ in the clouds, and hence, is inversely proportional to the cloud cover area $\Pi_{cloud}$. Then the following approximate equality can be written:

$$\theta = \frac{\langle \Delta V_w \rangle}{\langle V_w \rangle} \sim k_\Pi \frac{\Pi_{t=0}}{\Pi_t} \sim k_\Phi \frac{\Phi_{t=0}}{\Phi_t} = \Phi_\oplus^{-1}, \qquad (11)$$

where $\Pi_t / \Pi_{t=0} = \Pi_\oplus$ and $\Phi_t / \Phi_{t=0} = \Phi_\oplus$ are the variations of the cloud cover area $\Pi_t$ and the GCR intensity $\Phi_t$ at time $t$ with respect to analogous magnitudes $\Pi_{t=0}$ and $\Phi_{t=0}$ measured, for example, at the present point of time $t$=0; $k_\Pi$ and $k_\Phi$ are coefficients of proportionality. It is natural that within the framework of our one-zonal or, in other words, the point model of climate the GCR intensity $\Phi_t$ is latitude-averaged.

The Eqs. (10) and (11) require supplementary comment. It will be recalled that the results of Refs. (Marsh and Svenmark, 2000; Svensmark and Calder, 2007) show the existence of correlation between GCR intensity and the degree of cloud cover. At the same time, detailed analysis of experimental IPCC-data of 22-year solar cycle (Sloan and Wolfendale 2008) shows that only 23% of cloud cover can be caused by the cosmic rays at best, or more precisely, due to so-called ion-induced mechanism of the CCN formation in the lower atmosphere. Moreover, Sloan and Wolfendale (2008) logically come to conclusion that, if the correlation between GCR intensity and the degree of cloud cover actually takes place, more than 77% of cloud cover must be caused by the unknown CCN source, whose mechanism of nucleation is differs from ion-induced mechanism, but whose intensity is also modulated by the solar activity.

Surprisingly, but, firstly, such a "neutral" mechanism was theoretically predicted for a long time (Kulmala, 2003), and, secondly, it was recently verified by direct experiments on atmospheric nucleation (Kulmala et al, 2007). One of main conclusions of this paper consists in the fact that the so-called "neutral" mechanism of aerosols nucleation dominates over the ion-induced mechanism at least in boreal forest condition (Kulmala et al, 2007).

On other hand, it is known that the main source of CCN above oceans is dimethylsulphide (DMS), which is produced by the plankton in a sea water and oxidizes in the atmosphere to neutral sulphate aerosol (Charlson et al, 1987). In this case, the biological regulation of climate takes place due to feedback (by way of the DMS production) as a result of insolation and temperature forcing on phytoplankton population. At the same time, the DMS production rate is modulated by insolation, and the average annual part of cloud cover of the Earth conditioned by such a DMS-source is more than 40% (Charlson et al, 1987). Along with possible similar sources, the DMS-source is one of the main candidates to role of additional source of neutral CCN, which in combination with the source of ion-induced nucleation mechanism (GCR) can ensure the real correlation between the GCR intensity and degree of cloud cover.

So, Eqs. (10)-(11) have the following physical sense. It is possible to assume that the dependence (6) ($<V_w> \cong a_w + b_w T$) emulates the neutral mechanism of nucleation, while the dependence (11) emulates the ion-induced mechanism of nucleation. In other words, when the cosmic ray intensity is "low" and the ECS temperature grows, it means that the neutral mechanism of nucleation, which ensures the growth of cloud volume (and re-emission energy, respectively) up to some threshold value, prevails. And vice versa, when the cosmic ray intensity is "high", the cloud volume produced in this case due to the competition of the neutral and ion-induced mechanisms of nucleation also grows, but by a negligible margin. On the other words, it grows so that the total cloud volume and the corresponding area of cloud cover would exceed some threshold value, after which the cloud re-emission energy is sharply decreased. Obviously, that in this case the cosmic rays play the part of latent trigger starting a threshold mechanism of the variations of cloud re-emission power in the atmosphere.

To show evidently such a trigger effect we give the results of simulation of the simple submodels of climatic cloud-billiards in terrestrial electric field (Fig. 6), which takes into account the GCR-clouds effect on the basis of interdependent micro- and macrophysics of so-called indirect aerosol effects (Twomey, 1977, 1991; Kaufman and Fraser, 1997; Breon et al, 2002; Kaufman et al, 2002; Lyon, 2000).

For simplicity we consider here only two extreme cases. The initial concentration of condensation nuclei in cloud-billiards is equal in both cases. The first case (Fig. 6a) is characterized by some average intensity of GCR which, according to well-known estimation [3], generate no more than 23% of "charged" condensation nuclei of total number of all condensation nuclei (neutral and charged) due to ion-induced mechanism. In the second case (Fig. 6b) the GCR intensity is on the average 15 % down from the first extreme case according to the experimental data of Refs. (Marsh and Svensmark, 2000; Svensmark and Calder, 2007). Let us consider these scenarios in more detail.

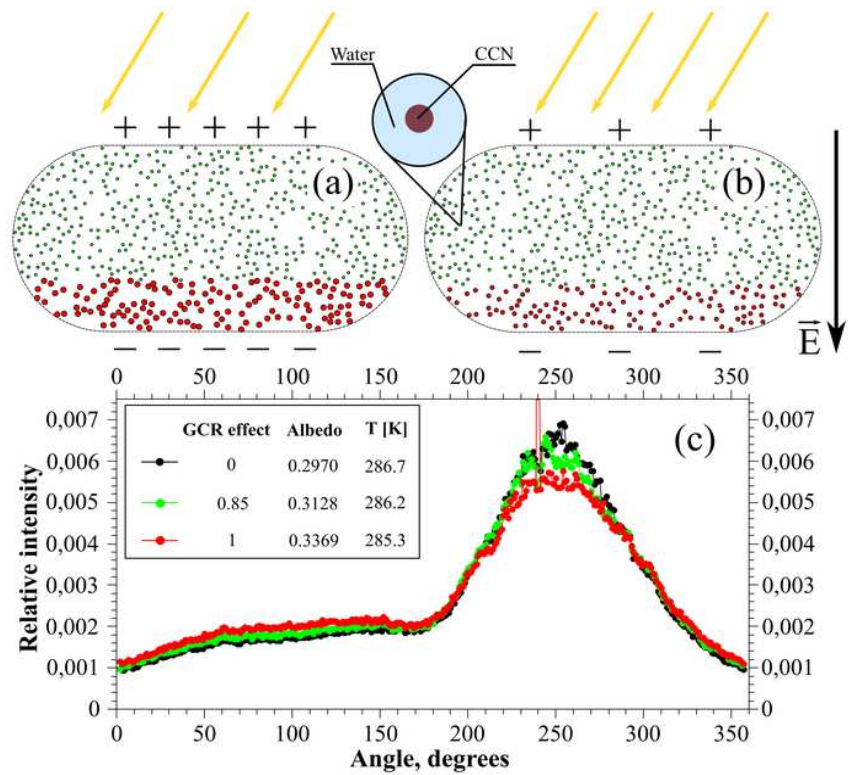

Fig. 6. Schematic drawing and simulation of the climatic billiard model emulating the GCR-clouds effect:

(a) - a cloud billiards corresponding to the first extreme case, when the GCR intensity is maximum and solar radiation intensity is minimum,

(b) a cloud billiards corresponding to the second extreme case, when the GCR intensity is minimum and solar radiation intensity is maximum. A cloud billiards is in the electric field ($\vec{E}$) and consists of droplets produced due to ion-induced mechanism (grey circles) and neutral mechanism (hollow circles).

(c) the relative intensity of solar radiation scattered in forward and backward semisphere depending on the scattering angle for different GCR intensities. The temperature $T$ is calculated by Eq. (19).

*The first case* (Fig. 6a). The following parameters of billiards were used for the simulation: the radius of half-circle is 136 arbitrary units, the center distance of half-circles is 299 arbitrary units, the radiuses of neutral and charged droplet[3] are 1 and 2 arbitrary units, respectively, the radiuses of inner condensation centre of the neutral and charged droplet are 0.1 and 0.2 arbitrary unit (justification of choice of the radius of internal condensation centre is given below). The total number of droplets of water is 600. To generate the random coordinates of each droplet of water we use the random number generator so that the distance between droplets is greater than the double radius of charge droplet (grey circles in Fig. 6a), i.e., greater than 4 arbitrary units.

In other words, in the calculations we always consider the cloud-billiards of fixed volume filled by water vapour (index of refraction $n_v$=1.00) and by the "neutral" droplets of water ($n_w$=1.33), which are characterized by initial external and internal radiuses and the concentration $\mu$ =600 droplets/billiards area. Moreover, we assume that

− billiards is under the action of solar radiation and cosmic rays characterized by an averaged energy spectrum and constant intensity;

− the solar radiation on the boundary and inside of billiards surrounded by air ($n_{air}$=1) is scattered under the laws of geometrical optics (without taking into account an absorption in water vapour and water) and is totally reflected from the surface of internal "solid" condensation nucleus (see Fig. 6b);

− the inner radius of neutral and charged droplet is chosen so that in the second extreme experiment the albedo $\alpha$ or the relative integral intensity of solar radiation scattered in backward semisphere is $\alpha \cong 0.30$ according to well-known measurements of Goode at al. (2001);

− a cloud-billiards in the Earth's electric field has low conductivity (Carslaw et al, 2002) thereupon the charged particle diffusion is low in this field too.

During irradiation two key aerosol processes simultaneously take place in the billiards sensitive volume: charged CCN formation accompanied with condensation and further growth of charged droplets due to capture of neutral droplets of water (the second indirect aerosol effect), and growth of neutral droplets concentration due to concomitant process of cooling (the first indirect aerosol effect). Note that just competition between these two processes predetermines the real optical properties of cloud-billiards (Fig. 6). In more detail, it looks like this.

Under GCR and due to ion-induced mechanism of nucleation, respectively, charged CCN are formed in the volume of billiards, and after water vapour condensation on them the charged droplets of water are appeared. According to Refs. (Sloan and Wolfendale, 2008; Erlykin et al, 2008), their part is about 23% of the total number of droplets of water ("charged" and "neutral"). Then slowly moving in the gravity and electric fields of the Earth the positive charged droplets of water capture mainly the neutral droplets of water due to collisions and coalescence. At the same time, the inner and outer radiuses of charged droplet of water double (red circles in Fig. 6a), i.e., they become equal to 2 and 0.2 arbitrary units, respectively. A measure of increase of the droplet outer radius at qualitative level is predetermined by behavior of the experimental dependence of CDR on AI (Fig. 2).

---

[3] Here we understand by droplet some aggregated ensemble of large number of real droplets, whose quantity can be easily determined from geometric considerations. Such formulation of the problem with an allowance for some assumptions (see below) and rules of geometrical optics is adequate to a computing experiment in real geometry.

The additional volume of water in the form of the charged droplets leads to increase of solar radiation albedo. On the other hand, the cooling of cloud-billiards caused by increase of albedo (the Twomey effect) stimulates the water vapour condensation and, consequently, results in the increase of neutral droplet concentration.

Let us consider now kinetics of negative charged droplets. Even in their minor upward motion under the Earth electric field, they attach to itself the nearest-neighbor "sessile" neutral droplets with the sufficiently high probability. A moment, when gravity becomes higher than the Coulomb force, comes quickly, and droplets begin slowly to move in the opposite direction, i.e. downward. After that, they replicate the qualitative behavior of the positive charged droplets or coalesce with the positive charged droplets transforming into neutral drops.

In the end, close to the cloud base the charged droplets reach an equilibrium charge configuration in the form of the so-called double electrical layer (DEL), where the negative and positive charged droplets are located in the top and bottom, respectively (Fig. 6 a-b). At the same time, the opposite directivity of the DEL intensity and the Earth electric field intensity predetermines low conductivity inside of DEL. Note that the existence of low conductivity regions in clouds is known for a long time (Carslaw et al., 2002). Apparently, just such a double electrical layer formation close to the cloud base is physical reason of the sufficiently high stability of cloud, which, in natural way, manifests in the increase of content of water in the cloud, cloud lifetime and, consequently, cloud cover area. Thus, within the framework of our model just a mechanism of the double electrical layer formation close to the cloud base is that reason which explains micro- and macrophysics of the so-called second indirect aerosol effect.

As mentioned above, competition of these two indirect aerosol processes (increase of the volume of charged droplets of water and production of the new neutral droplets of water) predetermines the proper optical properties of cloud-billiards. It will be recalled that in our model increase of the volume of charged droplets of water takes place mainly due to neutral droplets. In the calculation, we assume that due to competition the total number of charged droplets in billiards is approximately 23% (i.e., 140 droplets) of the total number of droplets (600 droplet/billiards). Formally, it means that charged droplets in process of growth "ate up" volume of water proportionate to the volume of all neutral droplets induced by the Twomey effect.

*The second case* (Fig. 6b). According to experimental data (Marsh and Svensmark, 2000; Svensmark and Calder, 2007), the solar radiation intensity is 0.1 % up on the first extreme case. At the same time, the cosmic ray intensity is 15 % down from the first extreme case. This corresponds to 119 "charged" droplets of the total number of droplets of water (600 droplet/billiards). The inner and outer radiuses of charged droplet are 1.5 and 0.15 arbitrary units, respectively. The decrease of these radiuses relative to the first extreme case emulates decrease of the Earth electric field intensity or, more precisely, fair-weather conduction current (about 2 picoamps per square meter) flowing between the ionosphere and Earth, which, according to Carslaw et al. (2002), is modulated by GCR ionization.

The results of computing experiment are presented in Fig. 6c, where one can see the pronounced trigger character of the change of sun radiation intensity scattered in forward and backward semispheres. A scattering angle is counted in the coordinate system whose center is coincided with the geometrical center of billiards. To simulate sunbeam trajectories incident on the billiard surface at an angle of 60 degrees the rules of geometrical optics were used. According to the experimental estimation of insolation variation (Marsh and Svensmark, 2000; Svensmark and Calder, 2007), the number of trajectories in the first and second experiments differs by 0.1% and is equal to 999000 and 1000000, respectively. At the same time, the variation of relative integral intensities of solar radiation forward scattered (Fig. 6c) is ~ 2.41 % in both cases, that with an allowance for solar constant 1366.2 W/м$^2$ (Frohlich and Lean, 1998) leads to the deficit of forward scattered radiation about 32.9 W/m$^2$. This is in good agreement with the current climatic estimate for the net forcing of the global cloud cover ~ 27.7 W/m$^2$ cooling (Marsh and Svensmark, 2000; Svensmark and Calder, 2007).

Note that this variation or, in other words, the deficit of forward scattered radiation can be refined if in the simulation to take into account absorption effects and small changes (~2%) of cloud cover due to diffusion of charged droplets (the first extreme case) in the Earth electric field. However, it is not question of principle because the presented semi-phenomenological example can be examined only as the visual illustration of trigger effect due to indirect aerosol effects against the background of intense competition between the neutral and ion-induced mechanisms of stable condensation nucleus formation in the Earth electric field. It is caused by the fact that description of coalescence of "stationary" neutral droplets and charged droplet, which slowly diffuses in the Earth electric field, requires high computational resources and a separate examination. However, this goes beyond the limits of the present work.

It is clear that such threshold mechanism of variations of cloud re-emission power operates effectively only when the relatively small variations of cloud volume and the corresponding area of cloud cover (which is experimentally observed (Marsh and Svenmark, 2000; Svensmark and Calder, 2007; Stozhkov, 2003; Usoskin et al, 2004)) take place just in the area of bifurcation point of cloud re-emission power (with respect to their volume or coverage area). It is confirmed by the bistability of some processes in the ECS (in particular, of cloud re-emission), which are "genetically" connected with well-known bistability of solar activity. Therefore, such bifurcation mechanism of variations of cloud re-emission power, which reflects the sense of physical nature of correlation between the cosmic ray intensity and cloud cover, is putted in Eqs. (10) and (11).

Further, we suppose that the average radiation power of the liquid water unit mass $\varepsilon_w$ in the first approximation depends linearly on ECS temperature. Then taking into account Eqs. (6) and (11) we introduce the linear dependence of $\varepsilon_w$ on the ECS temperature in Eq. (10):

$$G_w(T,t) = \frac{\gamma h}{\langle V_{atm} \rangle} \varepsilon_w \rho_w \theta \langle V_w \rangle = \frac{\gamma h}{\langle V_{atm} \rangle} \rho_w \left( a_{w\varepsilon} T^2 + b_{w\varepsilon} T + c_{w\varepsilon} \right) \Phi_{\oplus}^{-1}(t), \qquad (12)$$

where $h$ is the average height of the atmosphere $\langle V_{atm} \rangle \approx \gamma h$.

Let us consider now the details and difficulties in the calculation of time dependence of cosmic ray intensity. First, when we calculate the intensity, it is has necessary to take into account the two influencing factors in atmosphere upper layers:

a) modulations caused by the solar wind (this effect correlates with the solar activity and has strong temporal dependence at annual time scale, as well as it correlates with the Earth orbital eccentricity at the time scale about $10^4$ years, which is the important case for us);

b) cutoff of the low-energy part of cosmic ray spectrum owing to the geomagnetic fields (this effect depends on the locality latitude and weakly depends on time).

Second, solar wind "decelerates" the cosmic rays. This effect is usually described by the diffusion convection model, which gives the following formula for the observed spectrum $I(p,r,t)$ (Parker, 1963):

$$\Phi_\oplus(p,r,t) = \frac{I(p,r,t)}{I(p)} = \exp\left[-\int_{r_{min}}^{r_{max}} \frac{\upsilon(t)}{D(p,r',t)} dr'\right], \qquad (13)$$

where $r_{min}$ is the distance from the Earth, $r_{max}$ is the distance of solar wind from the Sun, $\upsilon(t)$ is the solar wind velocity, $D$ is diffusion coefficient, $I(p)$ is the spectrum in interstellar space depending on the particle momentum $p$.

This effect is maximal both at maximum solar activity at time scales $\geq 10$ years and at the minimal eccentricity of the Earth's orbit at the millennial time scale. For example, when solar activity is minimal, the proton beam with energy of 1 GeV is twice as much then at maximal solar activity. According to experimental data (Svensmark and Friis-Christensen,1997; Svensmark, 1998), this effect is decreased to <10 % at $E_p$=10 GeV.

Unfortunately, calculation by Eq. (13) or the use, for example, of the geophysical reconstruction data (by tracks in meteorites) for restoration of the temporal evolution of GCR intensity at the millennial time scales is now impossible. For this reason, it seems that the verification of global climate model with two governing parameters (insolation and cosmic ray intensity) by comparing the solution of model with the known experimental time series of palaeotemperature (e.g., the Vostok ice core data (Petit et al, 1999) over the past 420 kyr and the EPICA ice core data (EPICA community members, 2004) over the past 740 kyr) is, at first glance, undecidable. In our opinion, however, there is one "clue", which allows to evade this problem, although approximately and with certain limitations. This is done in the following manner.

As is known, the intensity of GCR, which reach the midtroposphere, is modulated by the solar wind. However, when the solar wind "sweeps" more or less the galactic protons, it perturbs to one extent or another, the Earth magnetic field due to the magnetic reconnection (Priest and Forbes, 2000; Coley, 1985). If these physically determined connections translate into language of formulas, we obtain the following approximate inverse dependence between the relative variations of GCR intensity $\Phi_\oplus$ and the relative variations of the Earth's magnetic field $H_\oplus$:

$$\Phi_\oplus(t) = k_\Phi \frac{\Phi_t}{\Phi_{t=0}} \sim k_H \frac{H_{t=0}}{H_t} = k_H H_\oplus^{-1}(t), \quad H_\oplus(t) \geq 0.5, \qquad (14)$$

where $H_t/H_{t=0} = H_\oplus$ is the relative variation of the Earth magnetic field $H_t$ in time $t$ with respect to analogous magnitude measured at this moment in time $t$=0.

It will be recalled once more that Eq.(14) represents rough approximation, and the physical sensce of its limitation ($H_\oplus \geq 0.5$) will be considered below. Our desire to turn from the time sample of $\Phi_\oplus$ to the analogical sample of $H_\oplus$ by Eq. (14) is explained by the fact that the time sample of $H_\oplus$ can be determined from the experimental magnetic palaeodata (Yamasaki and Oda, 2002). For example, the time evolution of the relative variations of Earth magnetic field $H_\oplus(t)$ at the millennial time scale can be calculated by the following expression

$$H_\oplus(t) = \frac{H_t}{H_{t=0}} = \frac{M_t \chi_0}{\chi_t M_0},\qquad(15)$$

where $M = \chi H$ is magnetic moment per volume unit or magnetization; $\chi$ is magnetic susceptibility (Butler, 1998). The necessary experimental data of the magnetization $M_t$ and magnetic susceptibility $\chi_t$ at the millennial time scale for time $t \in [0, 2.25]$ million years can be found in Ref. (Yamasaki and Oda, 2002).

In this way, taking into account Eq.(14), the expression (12) for re-emission energy of liquid water in clouds looks like:

$$G_w(T,t) = \frac{\gamma h}{\langle V_{atm}\rangle}\rho_w\left(a_{w\varepsilon}T^2 + b_{w\varepsilon}T + c_{w\varepsilon}\right)k_H H_\oplus(t),\qquad(16)$$

The expression for the heat energy power $G_v(T,t)$ reradiated by water vapour in the clouds can be obtained in analogous way:

$$G_v(T,t) = \frac{\gamma h}{\langle V_{atm}\rangle}\rho_v\left(a_{v\varepsilon}T^2 + b_{v\varepsilon}T + c_{v\varepsilon}\right)k_H H_\oplus(t),\qquad(17)$$

where $\varepsilon_v$ is the average radiation power of water vapour per mass unit, $Wkg^{-1}$; $\rho_v$ is water vapour density; $\langle V_{atm}\rangle \approx \gamma h$.

To determine the functional dependence of the heat energy power $G_{CO_2}(T,t)$ on the ECS temperature we use the known experimental data of surface paleotemperature evolution and carbon dioxide concentration over the past 420 kyr obtained from ice core at the Antarctic station "Vostok" (Petit et al, 1999) and over the past 740 kyr from the EPICA ice core (EPICA community members, 2004). It is obvious (EPICA community members 2004) that these data have strong linear correlation. Therefore, we can suppose that the dependence of the heat energy power $G_{CO_2}(T,t)$ on the ECS temperature is also linear:

$$G_{CO_2}(T,t) = \frac{\gamma h}{\langle V_{atm}\rangle}\varepsilon_{CO_2}\beta T.\qquad(18)$$

where $\varepsilon_{CO_2}$ is the radiant energy of carbon dioxide per mass unit , $\beta$ is the accumulation rate of carbon dioxide in the atmosphere normalized by unit of temperature, $kgK^{-1}$.

Theoretically this dependence can be also explained within the three-mode model of radiation kinetics in the atmosphere (Glushkov et al, 1996, 2000), which takes into account redistribution of energy and heat exchange in the atmospheric gas mixture $CO_2$-$N_2$-$O_2$-$H_2O$? which interact with electromagnetic radiation. Due to absorption of electromagnetic radiation by the molecules of atmospheric gases, a redistribution of molecules on the energy levels of internal degree of freedom takes place and absorption saturation leads to the changes of gas absorption factor. In our case, formation and accumulation of excited nitrogen molecules owing to the resonant transfer of excitation from the $CO_2$ molecules leads to the change of environment polarizability, but conserves the linear dependence of heat energy power on the ECS temperature (Glushkov et al, 1996, 2000).

Note that the dependence of the effective albedo on the ECS averaged temperature is chosen in the form of the following continuous parameterization

$$\alpha = \alpha_0 - \eta_\alpha \cdot (T - 273). \tag{19}$$

Eq. (19) describes well, for example, the albedo behaviour (at $\alpha_0$=0.7012, $\eta_\alpha$=0,0295 $K^{-1}$) in the temperature range 282-290 K.

Finally, collecting all partial contributions of heat flows (16)-(19) and $I_{Earth} = \gamma \delta(\sigma T^4)$ into the resulting energy-balance Eq. (9), we have

$$U^*(T,t) = \frac{1}{4}T^4 + \frac{1}{2}a(t) \cdot T^2 + b(t) \cdot T, \tag{20}$$

where

$$a(t) = -\frac{1}{4\delta\sigma}a_\mu H_\oplus(t), \tag{21}$$

$$b(t) = -\frac{1}{4\delta\sigma}\left[\frac{\eta_\alpha S_0}{4} + \frac{1}{2}\beta + \frac{1}{2}b_\mu H_\oplus(t)\right], \tag{22}$$

$$U^*(T,t) = \frac{1}{4\delta\sigma}\left[\frac{1}{4}(1-\alpha_0-273\eta_\alpha)S_0 + \frac{1}{2}b_{CO_2}\,h/\langle V_{atm}\rangle + \frac{1}{2}c_\mu H_\oplus - U(T,t)\right], \tag{23}$$

$$a_\mu = k_H(\rho_w a_{w\varepsilon} + \rho_v a_{w\varepsilon})\gamma^{-1}, \text{ Wm}^{-2}\text{K}^{-1}$$

$$b_\mu = k_H(\rho_w b_{w\varepsilon} + \rho_v b_{w\varepsilon})\gamma^{-1}, \text{ Wm}^{-2}\text{K}^{-1}$$

$$c_\mu = (\rho_w c_{w\varepsilon} + \rho_v c_{w\varepsilon})\gamma^{-1}, \text{ Wm}^{-2},$$

where $a_{w\varepsilon}$, $a_{v\varepsilon}$, $b_{w\varepsilon}$, $b_{v\varepsilon}$, $c_{w\varepsilon}$, $c_{v\varepsilon}$ are constants, whose dimensions are determined by Eqs. (16) and (17), respectively.

It is obvious that Eq. (20) describes the collection of energy-balance functions $U^*(T,a,b)$, which depend on two governing parameters $a(t)$ and $b(t)$. Also, this collection represents so-called assembly-type catastrophe potential (Gilmore, 1985).

In the sequel, we will be interested in the form of "disturbed" equation (20) or, more precisely, the form of assembly-type catastrophe equation (20) with respect to the increment $\Delta T = T - T_0$ of the following type: $U(T_0 + \Delta T, a, b) - U(T_0, a, b) = \Delta U$, where $T_0$ is the average temperature of ECS averaged on the corresponding time span $\Delta t$. At the same time, an increment for the first term on right-hand side of Eq. (20) was used in the following equivalent form:

$$(T_0 + \Delta T)^4 - T_0^4 \cong 7 \cdot 10^{-3} \cdot T_0^3 \cdot (\Delta T)^4 + 4 \cdot T_0^3 \cdot \Delta T, \quad for \quad \Delta T = 0 \div 4K, \tag{24}$$

where the approximation average error in given temperature range does not exceed 0.01%.

Note that the insolation normalized variation

$$\Delta W = \frac{W - \langle W_0\rangle}{\sigma_S} \tag{25}$$

with the average <$\Delta W$>=0 and dispersion $\sigma_{\Delta W}^2$ =1 is used more often for the ECS simulation.

Designing an equation of Eq. (20) type with respect to $\Delta T$, we obtain the following expression for increment of the heat power $\Delta U^*$:

$$\Delta U^*(\Delta T, t) = \frac{1}{4}\Delta T^4 + \frac{1}{2}\tilde{a}(t)\cdot\Delta T^2 + \tilde{b}(t)\cdot\Delta T,$$ (26)

where

$$\tilde{a}(t) = -\frac{37.6}{\sigma T_t^3}a_\mu H_\oplus(t) = -\tilde{a}_0\cdot H_\oplus(t),$$ (27)

$$\tilde{b}(t) = -\frac{37.6}{\sigma T_t^3}\left[\eta_\alpha\frac{S_0 + \Delta\hat{W}(t)\sigma_S}{4} - 4\delta\sigma T_t^3 + \frac{1}{2}\beta + \frac{1}{2}(2a_\mu T_t + b_\mu)H_\oplus(t)\right] =$$

$$= -\tilde{b}_0\left[\eta_\alpha W_{reduced}(t) - 4\delta\sigma T_t^3 + \frac{1}{2}\beta + \frac{1}{2}(2a_\mu T_t + b_\mu)H_\oplus(t)\right],$$ (28)

$S_0$ is "solar constant"; $\Delta\hat{W}(t)$ is the reduced normalized dispersion of insolation; $\sigma_S$ is reduced roof-mean-square deviation; $4W_{reduced} = S_0 + \Delta\hat{W}(t)\,\sigma_S$ is reduced average annual insolation of the Earth. The substantiation of the choice of reduced values $\Delta\hat{W}(t)$, $\sigma_S$ and $W_{reduced}$ will be considered in more detail in part 2 of present paper.

Finally, we give the canonical form of assembly-type catastrophe variety, which represents a set of points $(\Delta T, \tilde{a}, \tilde{b})$, satisfying the following system of equations:

$$\frac{\partial}{\partial T}U^*(T, t) = T_t^3 + a(t)\cdot T_t + b(t) = 0.$$ (29)

$$\frac{\partial}{\partial(\Delta T)}\Delta U^*(\Delta T, t) \cong \Delta T_t^3 + \tilde{a}(t)\cdot\Delta T_t + \tilde{b}(t) = 0.$$ (30)

Thus, the general bifurcation problem of finding the solution $T(t)$ and perturbed solution $\Delta T(t)$, which correspond to the average global temperature and an increment, is reduced to the finding of solution set of Eqs. (29)-(30) for the appropriate joint trajectory $\{a(t), b(t)\}$ and $\{\tilde{a}(t), \tilde{b}(t)\}$ in the space of governing parameters (Fig. 7).

## 4. Principle of structural invariance of hierarchical models

Before to discuss the details of computational experiment and its results (Rusov et al, 2010), let us pay attention to some conceptually important moments related to the hierarchy of climate states on the different time scales and possible links between them. In this sense hidden connection between the multi-zonal model of Earth weather on the decennial time scale and one-zonal model of global climate on the millennial time scale is most interesting in respect to physics of hierarchy of climate states on the different scales of time there. Strangely enough, such a connection exists, for example, due to principle of structural invariance of hieratic equations. Let us consider the features of this principle.

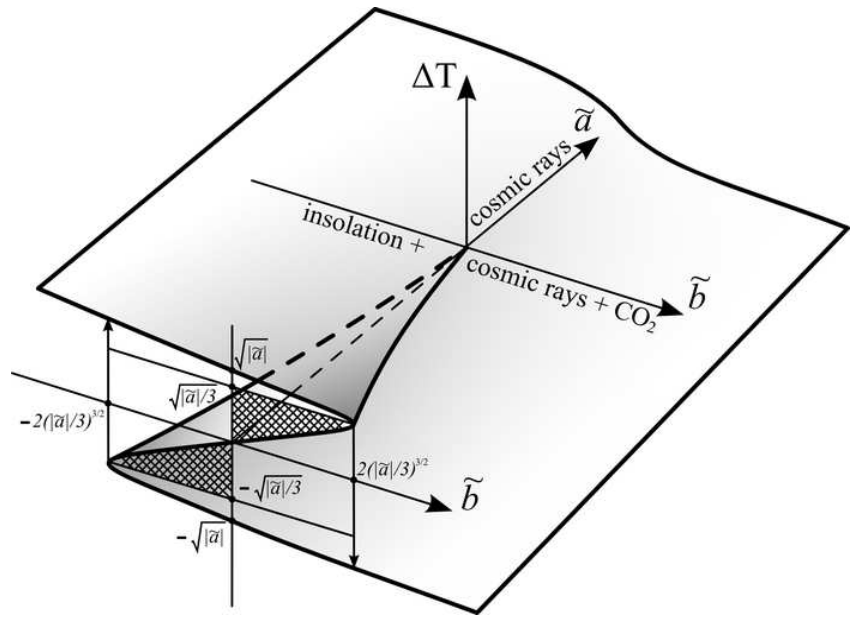

Fig. 7. Canonical form of the variety of assembly-type catastrophe as a set of points $(\Delta T, \tilde{a}, \tilde{b})$ satisfying Eq. (30) type. Crosshatched areas are the regions of the instable solutions of Eq. (30).

Toward this end let us to put question: "What is in store for us: the global warming due to non-controlled growth of the anthropogenic $CO_2$ concentration in the atmosphere or, for example, the sharp global cooling due to the minimal (for the last two centuries) solar activity starting from 2020 year (the so-called Schwabe solar cycle)?" To answer this question it is necessary to know the real physical mechanism of climate change formation on short time scale (for example, 20-30 years) or, more precisely, the real physical model of Earth weather. One can state that none of the living now weather multizonal models can answer this question, because all these models do not satisfy the main and, certainly, key principle of hierarchical model construction, which consists in the structural invariance of balance equations of global climate and global weather. It means that the system of equations of weather multizonal model convoluted into the balance equation of one-zonal model practically fully keeps the structure and properties (governing parameters) of the global climate model on the intermediate (millennial) and long (millionth) time scale. In other words, the principle of structural invariance of the equations of global climate and global weather predetermines and, in that way, sets the unambiguous and holistic strategy for study of nonlinear physics of global climatic changes evolving on the different time scales.

Essence of such a strategy realized within the framework of presented paper and Ref. (Rusov et al, 2010) is simple. At first one has to research and to found the group of governing parameters, on which the energy-balance equation of global climate model depends on intermediate (millennial) time scale. Afterwards one has to verify the model (Rusov et al, 2010) by comparison with the well-known experimental paleo-temperature data (for example, (EPICA community members, 2004) and (Petit et al, 1999)). In the case of good agreement between theory and experiment, it means that the model correctly "guesses" the physical sense of governing parameters (for example, the variations of insolation and cosmic ray intensity). Only now that physics of process is "guessed", it is possible, using the found model of global climate (on the millennial time scale), to begin the multizonal model of Earth weather construction (on a short time scale) taking into account the peculiarities of indicated time scales.

Part 2 of this paper (Rusov et al, 2010) is devoted to verification of the solution of Eqs. (29)-(30) of energy-balance model of the Earth global climate based on the well-known experimental data of Earth surface palaeotemperature evolution , i.e. the Vostok ice core data over the past 420 kyr and the EPICA ice core data over the past 740 kyr.